\documentstyle[sprocl,epsf,axodraw]{article}

\newcommand{\beq}{\begin{equation}}
\newcommand{\eeq}{\end{equation}}
\newcommand{\bqa}{\begin{eqnarray}}
\newcommand{\eqa}{\end{eqnarray}}

\begin{document}

\renewcommand{\theequation}{\thesection.\arabic{equation}}

\title{Three Loop Free Energy Using Screened Perturbation Theory
\footnote{ Talk given at Conference on Strong and Electroweak
Matter (SEWM 2000), Marseille, France, 14-17 June 2000.}}

\author{Jens O. Andersen}

\address{Physics Department, Ohio State University, Columbus OH 43210, USA}

\maketitle\abstracts{The conventional weak-coupling expansion
for the pressure of a hot plasma shows no sign of convergence unless the
the coupling constant $g$ is tiny.
In this talk, I discuss screened perturbation theory (SPT)
which is a reorganization of the perturbative expansion by adding and 
subtracting a local mass term in the Lagrangian.
We consider several different mass prescriptions, and compute the
pressure to three-loop order. 
The SPT-improved approximations appear to converge for rather large values of
the coupling constant.
}


\section{Introduction} 
The heavy-ion collision experiments at RHIC and LHC give us for the first time
the possibility to study the properties of the high-temperature phase
of QCD. There are many methods that can be used to calculate the
properties of the quark-gluon plasma. One of these methods is lattice gauge
theory, which gives reliable results for equilibrium properties such as the
pressure but cannot easily be applied to real-time processes.
Another method is the weak-coupling expansion, which can be applied to
both static and dynamical quantities.
However, it turns out that the weak-coupling expansion for e.g. the pressure
does not converge unless the strong coupling constant $\alpha_s$ is tiny.
This corresponds to a
temperature which is several orders of magnitude 
larger than those relevant for experiments at RHIC and LHC.

\begin{figure}[htb]
\epsfysize=5cm
\centerline{\epsffile{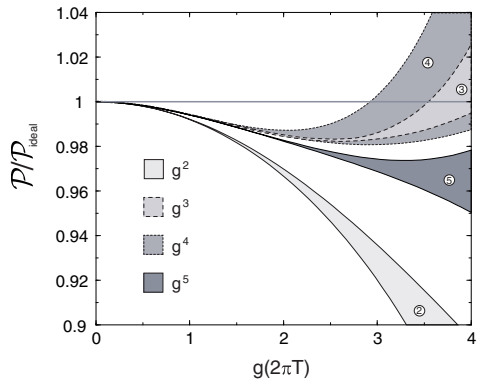}}
\caption[a]{Weak-coupling expansion to orders $g^2$, $g^3$, $g^4$, and $g^5$
for the pressure normalized to that of an ideal gas as a function
of $g(2\pi T)$.}
\label{fpert}
\end{figure}

The poor convergence of weak-coupling expansion also shows up in the 
case of scalar field theory.
For a massless scalar field theory with a $g^2\phi^4/4!$ interaction, 
the weak-coupling expansion for the pressure to 
order $g^5$ is~\cite{arnold-zhai}$^{\!-}$\cite{Braaten-Nieto:scalar}
\bqa\nonumber
{\cal P} &=& {\cal P}_{\rm ideal} \left[
1-{5\over4}\alpha+{5\sqrt{6}\over3}\alpha^{3/2}+{15\over4}
\left(\log{\mu\over2\pi T}+0.40\right)\alpha^2\right.\\ 
&&\left.-{15\sqrt{6}\over2}\left(\log{\mu\over2\pi T}-{2\over3}\log\alpha
-0.72\right)\alpha^{5/2}+{\cal O}(\alpha^3\log\alpha)\right]\;,
\eqa
where ${\cal P}_{\rm ideal} = (\pi^2/90)T^4$
is the pressure of the ideal gas of a free massless boson,
$\alpha=g^2(\mu)/(4\pi)^2$, and $g(\mu)$ is the 
$\overline{\rm MS}$ coupling constant at the renormalization scale $\mu$.
In Fig.~\ref{fpert}, we show the successive perturbative approximations to
${\cal P}/{\cal P}_{\rm ideal}$ as a function of $g(2\pi T)$. Each partial
sum is shown as an error band obtained by varying $\mu$ 
from $\pi T$ to $4\pi T$.
To express $g(\mu)$ in terms of
$g(2\pi T)$, we use the numerical 
solution to the renormalization group equation
$\mu{\partial\over \partial\mu}\alpha=\beta(\alpha)$ 
with a five-loop beta  function.
The lack of convergence of the perturbative series is evident in 
Fig.~\ref{fpert}. The band obtained by varying $\mu$ by a factor of two
is a lower bound on the theoretcal error involved in the calculations. Another
indicator of the error is the difference between successive approximations.
From Fig.~\ref{fpert}, we conclude that the error grows quickly for $g\geq1.5$.

\section{Screened Perturbation Theory}

Screened perturbation theory, which was introduced by 
Karsch, Patk\'os and Petreczky~\cite{K-P-P}, is simply a reorganization of the perturbation
series for thermal field theory.

The Lagrangian density for a massless scalar field with a $\phi^4$
interaction is
\bqa
\label{l1}
{\cal L}={1\over2}\partial_{\mu}\phi\partial^{\mu}\phi
-{1\over 24}g^2\phi^4+\Delta{\cal L}\;,
\label{barel}
\eqa
where $g$ is the coupling constant and $\Delta{\cal L}$ includes counterterms.
The Lagrangian
density is written as
\bqa
\label{SPT}
{\cal L}_{\rm SPT}=-{\cal E}_0 + {1 \over 2} \partial_\mu\phi\partial^\mu\phi
	-{1\over2}(m^2-m_1^2)\phi^2 - {1\over24} g^2 \phi^4
	+\Delta{\cal L}
	+\Delta{\cal L}_{\rm SPT}\;,
\eqa
where ${\cal E}_0$ is a vacuum energy density
parameter and we have added and subtracted
mass terms.
If we set ${\cal E}_0=0$ and $m_1^2=m^2$, we recover the original 
Lagrangian~(\ref{barel}).
Screened perturbation theory is defined by taking $m^2$ to be of order 
$g^0$ and $m_1^2$ to be of order $g^2$, expanding systematically in powers
of $g^2$, and setting $m_1^2=m^2$ at the end of the calculation.
This defines a reorganization of perturbation theory in which the expansion
is around the free field theory defined by
\bqa
\label{freesca}
{\cal L}_{\rm free}=-{\cal E}_0+{1\over2}\partial_{\mu}\phi\partial^{\mu}\phi
-{1\over2}m^2\phi^2\;.
\eqa
The interaction term is
\bqa
\label{scaint}
{\cal L}_{\rm int}= {-}{1\over24} g^2 \phi^4 + {1\over2}m_1^2\phi^2
+\Delta{\cal L}+\Delta{\cal L}_{\rm SPT}\;.
\eqa
At each order in screened perturbation theory,
the effects of the $m^2$ term in~(\ref{freesca}) are included to all orders.
However when we set $m_1^2 = m^2$, the dependence on $m$ 
is systematically subtracted out at higher
orders in perturbation theory by the $m_1^2$ term in~(\ref{scaint}).
At nonzero temperature, screened perturbation theory does not generate
any infrared divergences, because the mass parameter $m^2$ in the
free Lagrangian~(\ref{freesca}) provides an infrared cutoff. 
The resulting perturbative expansion is therefore a power series in $g^2$
and $m_1^2=m^2$ whose coefficients depend on the mass parameter $m$.

This reorganization of perturbation theory generates new ultraviolet
divergences, but they can be canceled by the additional counterterms
in $\Delta{\cal L}_{\rm SPT}$. The renormalizability of the Lagrangian 
in~(\ref{SPT}) guarantees that the only counterterms required
are proportional to 1, $\phi^2$, $\partial_{\mu}\phi\partial^{\mu}\phi$,
and $\phi^4$.
\subsection{Mass Prescriptions}
At this point I would like to emphasized that the mass parameter in SPT
is completely arbitrary, and we need a prescription for it.
The prescription of
Karsch, Patk\'os, and Petreczky for $m_*(T)$ is the solution to the
``one-loop gap equation'':
\bqa
\label{pet}
m_{*}^2=
4\alpha(\mu_*)\left[
\int_0^{\infty}dk\;{k^2\over\omega (e^{\beta\omega}-1)}
-{1\over8}\left(2\log{\mu_*\over m_*}+1\right)m_*^2
\right]\;,
\eqa
where $\omega=\sqrt{k^2+m^2_*}$ and $\alpha(\mu_*)=g^2(\mu_*)/(4\pi)^2$.
 
There are many posibilities for generalizing~(\ref{pet}) to 
higher orders in $g$.
Here, I consider three generalizations.
\begin{itemize}
\item{The {\it screening mass} $m_s$ is defined
by the location of the pole of the static propagator:}
\bqa
{\bf p}^2+m^2+\Pi(0,{\bf p})=0,\hspace{0.5cm} \mbox{at}\hspace{0.5cm}
{\bf p}^2=-m_s^2\;.
\eqa
\item{The {\it tadpole mass} $m_t$ is defined 
by the expectation value of $\phi^2$:}
\bqa
m_t^2=g^2\langle\phi^2\rangle\;.
\eqa
\item{The {\it variational mass $m_v$} is the solution to}
\bqa
{\!\!\!d\over dm^2}{\cal F}(T,g(\mu),m,m_1=m,\mu)=0\;
\eqa
Thus the dependence of the free energy on $m$ is minimized by $m_v$.
\end{itemize}
The above mass prescriptions all coincide at the one-loop order
and is given by~(\ref{pet}) above. The three masses differ
at the two-loop level and beyond. 
The two-loop gap equation for the tadpole mass 
turns out to be identical to the one-loop gap equation.
Note that the tadpole mass cannot
be generalized to gauge theories since the expectation value 
$\langle A_{\mu}A^{\mu}\rangle$ is a gauge-variant quantity.
Moreover, the screening mass in nonabelian gauge theories
is not defined beyond leading order in perturbation theory due to
a logarithmic infrared divergence.

\subsection{Results}
A thourough study of screened perturbation theory is presented in 
Ref.~\cite{spt}. Here, I only present a few selected results.

\begin{figure}[htb]
\epsfysize=5cm
\centerline{\epsffile{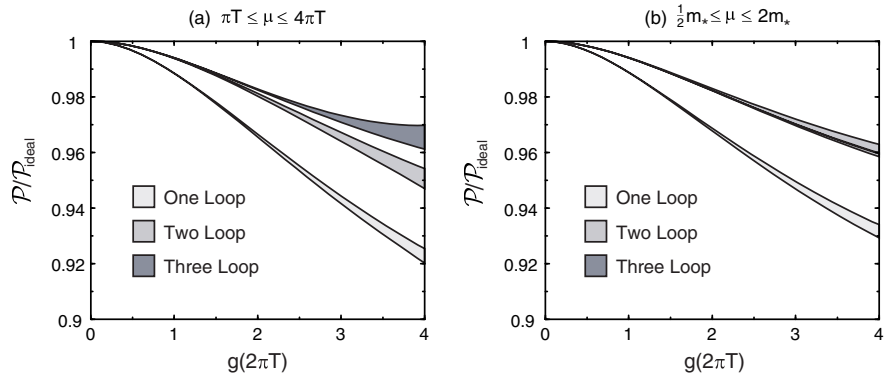}}
\caption[a]{One-, two-, and three-loop SPT-improved pressure
	as a function of $g(2\pi T)$ for
	(a) $\pi T < \mu < 4 \pi T$ and 
	(b) ${1 \over 2} m_* < \mu < 2 m_*$.
}
\label{fig7}
\vspace{-.5cm}
\end{figure}

In Fig.~\ref{fig7}, we show the one-, two- and three-loop SPT-improved
approximations to the pressure using the tadpole gap 
equation. 
The bands are obtained by varying $\mu$ by a factor of two around
the central values $\mu=2\pi T$ and $\mu=m_*$.
The choice $\mu=m_*$ gives smaller bands from varying the 
renormalization scale, but this is mainly due to the fact that $g(2\pi T)$
is larger than $g(m_*)$.
The vertical scale in Fig.~\ref{fig7}
has been expanded by a factor of about two compared to Fig.~\ref{fpert},
which shows the successive approximations using the weak-coupling expansion.  
All the bands in Fig.~\ref{fig7} lie 
within the $g^5$ band in Fig.~\ref{fpert}. 
Thus we see a dramatic improvement in the apparent convergence
compared to the weak-coupling expansion.
\section{Summary}
In this talk, I have briefly discussed SPT, which is a reorganization
of the perturbative expansion. In contrast to the weak-coupling expansion,
the SPT-improved approximations to pressure 
appear to converge for rather large values of the coupling constant.

Screened perturbation theory has been generalized to gauge theories
and is called hard-thermal-loop (HTL) perturbation theory~\cite{EJM1}. 
A one-loop calculation of the pressure with and without fermions has
already been carried out~\cite{EJM1}. Two-loop calculations are in 
progress~\cite{twoloophtl}

The fact that SPT shows very good convergence properties gives us hope
that HTL perturbation theory will be a consistent approach that can used
for calculating static and dynamical quantities of a quark-gluon plasma.
\section*{Acknowledgments}
This work was carried out in collaboration with Eric Braaten and Michael
Strickland.
The author would like to thank the organizers of SEWM 2000 for a stimulating meeting.
This work was supported in part by the U.~S. Department of
Energy Division of High Energy Physics
(grants DE-FG02-91-ER40690 and DE-FG03-97-ER41014)
and by a Faculty Development Grant
from the Physics Department of the Ohio State University.

\end{document}